# Tuning interfacial spins in antiferromagnetic / ferromagnetic / heavy metal heterostructures via spin-orbit torque

X. H. Liu[1,2], K. W. Edmonds[3], Z. P. Zhou[1,2], K. Y. Wang[1,2,4,5*]

[1]*State Key Laboratory for Superlattices and Microstructures, Institute of Semiconductors, Chinese Academy of Sciences, Beijing 100083, China*

[2] *Center of Materials Science and Optoelectronics Engineering, University of Chinese Academy of Sciences, Beijing 100049, China*

[3] *School of Physics and Astronomy, University of Nottingham, Nottingham NG7 2RD, United Kingdom*

[4] *Beijing Academy of Quantum Information Sciences, Beijing 100193, China*

[5]*Center for Excellence in Topological Quantum Computation, University of Chinese Academy of Science, Beijing 100049, China*

\* Corresponding e-mail: kywang@semi.ac.cn

Antiferromagnets are outstanding candidates for the next generation of spintronic applications, with great potential for downscaling and decreasing power consumption. Recently, the manipulation of bulk properties of antiferromagnets has been realized by several different approaches. However, the interfacial spin order of antiferromagnets is an important integral part of spintronic devices, thus the successful control of interfacial antiferromagnetic spins is urgently desired. Here, we report the high controllability of interfacial spins in antiferromagnetic / ferromagnetic / heavy metal heterostructure devices using spin-orbit torque (SOT) assisted by perpendicular or longitudinal magnetic fields. Switching of the interfacial spins from one to another direction through multiple intermediate states is demonstrated. The field-free SOT-induced switching of antiferromagnetic interfacial spins is also observed, which we attribute to the effective built-in out-of-plane field due to unequal upward and downward interfacial spin populations. Our work provides a precise way to modulate the interfacial spins at an antiferromagnet / ferromagnet interface via SOT, which will greatly promote innovative designs for next generation spintronic devices.

## I. INTRODUCTION

Antiferromagnets have numerous advantageous properties for future spintronics applications: robustness against external field, no stray fields, and ultrafast spin dynamics [1,2]. Especially, the recent discovery of electrical switching of an antiferromagnet by spin-orbit torque (SOT) shows that antiferromagnets can be electrically manipulated in similar ways to their ferromagnetic (FM) counterparts [3], stimulating considerable research in antiferromagnetic (AFM) spintronics [4-8]. To date, most work has focused on electrical manipulation of bulk properties of AFM materials [3-12]. Conversely, from the point of view of expanding the functionality and the design flexibility in AFM spintronic devices, developing methods to tune the interfacial properties of AFM materials through SOT is a vitally significant issue.

Exchange bias (EB) refers to a shift in the hysteresis loop along the magnetic field axis due to the interfacial exchange coupling between adjacent FM/AFM layers. This phenomenon has been extensively studied because of its technological importance, for example in read heads for magnetic storage or spin valves [13,14]. Moreover, it offers a unique tool to directly probe the AFM interfacial spin states and the interfacial exchange coupling. EB can be utilized to exert an internal effective field in a heavy metal (HM)/FM system to obtain deterministic SOT switching of a perpendicular magnetic anisotropy (PMA) magnetization [15-20]. In past decade, the electrical control of EB in FM/AFM heterostructures has been demonstrated using multiferroic AFM insulators YMnO$_3$, BiFeO$_3$, or Cr$_2$O$_3$ [21-23]. However, this effective electrical control faces a big challenge for metallic AFM materials, such as IrMn or PtMn. Very recently, Lin *et al.* discovered the concurrent switching of FM magnetization and EB by SOT in a HM/FM/AFM trilayer system [24].

Here, we report the high tunability of AFM interfacial spins by SOT combined with perpendicular or longitudinal magnetic fields in a HM/FM/AFM system. We can effectively switch the AFM interfacial spins between multiple different states, using different combinations of pulsed electrical currents and magnetic fields. Moreover, the field-free SOT induced switching of AFM interfacial spins is demonstrated. The realization of AFM interfacial multi-state spin switching via SOT with or even without external fields will enlarge the designability of AFM spintronics.

## II. EXPERIMENTAL

The stack structures of Ta (0.6)/Pt (3)/Co (0.8)/Ir$_{25}$Mn$_{75}$ ($t$)/Ta (2) (thickness in nanometers) with $t$ = 5, 6, 7, and 8 nm were deposited on thermally oxidized Si substrates by magnetron sputtering at room temperature. The bottom and top Ta layers were used for adhesion and capping layers, respectively. The base pressure was less than $1 \times 10^{-8}$ Torr before deposition, and the pressure of the sputtering chamber was 0.8 mTorr during deposition. No magnetic field was applied during the sputtering. The deposited rates for Ta, Pt, Co, and $Ir_{25}Mn_{75}$ films were controlled to be ≈ 0.016, 0.025, 0.012, and 0.015 nm/s, respectively. After that, the samples were patterned into Hall bar devices with channel widths of 10 μm by photolithography and Ar-ion etching. For field-annealing treatments, the fabricated devices were annealed at 250°C for 30 min at a base vacuum of $1 \times 10^{-7}$ Torr under out-of-plane [along **z** direction in Fig. 1(b)] magnetic field of 0.7 T, then were field-cooled to room temperature, by using oven for magnetic-field annealing (F800-35, East Changing Technologies, China). The Kerr characterization of magnetization hysteresis was taken using a NanoMoke3 magneto-optical Kerr magnetometer. The anomalous Hall effect measurements were carried out at room temperature with Keithley 2602 as the sourcemeter and Keithley 2182 as the nano-voltage meter.

### III. ANTIFERROMAGNETIC LAYER THICKNESS DEPENDENCE

Experiments were performed on Ta(0.6)/ Pt(3)/ Co(0.8)/ $Ir_{25}Mn_{75}$($t_{IrMn}$)/ Ta(2) (in nm) stacks, with $t_{IrMn}$ = 5, 6, 7 and 8 nm, as schematically illustrated in Fig. 1(a). Fig. 1(b) presents an optical micrograph of a typical Hall bar, along with the definition of the coordinate system. The initial anomalous Hall effect resistance ($R_H$) as a function of out-of-plane field ($B_z$) for samples with $t_{IrMn}$ = 5, 6, 7 and 8 nm are exhibited in Figs. 1(c)-(f). A square hysteresis loop was found for samples with $t_{IrMn}$ = 5 and 6 nm, with much larger coercivity for $t_{IrMn}$ = 6 nm, while no EB is observed for both samples. Similar results were also observed elsewhere [24]. Two-step switching behavior was observed for samples with $t_{IrMn}$ = 7 and 8 nm, with stronger out-of-plane pinning observed for $t_{IrMn}$ = 8nm [Fig. 1(e),(f)].

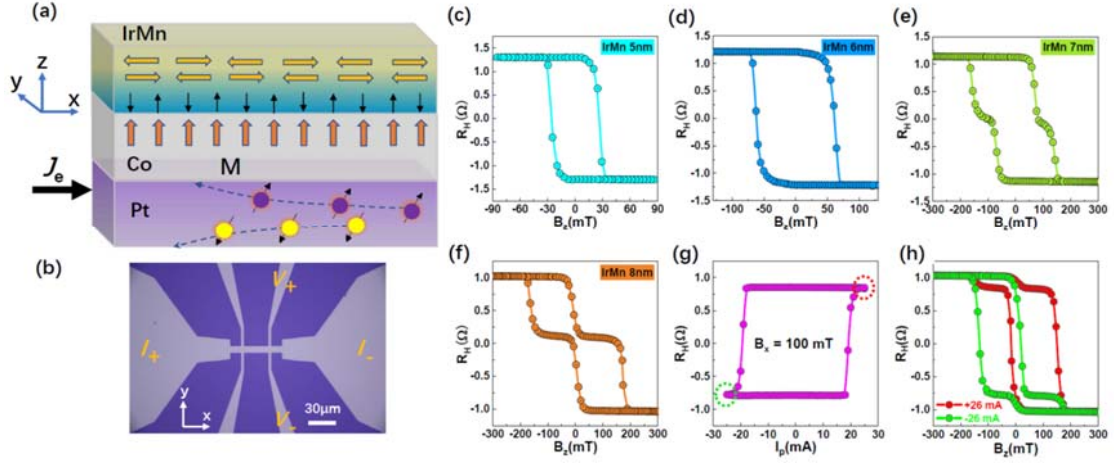

FIG. 1. Sample structure and magnetic properties. (a) Schematic of the studied HM/FM/AFM trilayer system with the definition of x-y-z coordinates. (b) Optical micrograph of the fabricated Hall device and measurement scheme. (c)-(f) Initial Hall resistance $R_H$ vs. perpendicular magnetic field $B_z$ curves for samples with $t_{IrMn}$ = 5, 6, 7 and 8 nm, respectively. The magnetic properties vary with $t_{IrMn}$, with a two-step behavior observed for samples with $t_{IrMn}$ = 7 and 8 nm. (g) $R_H$ versus current pulse amplitude $I_p$ under in-plane field $B_x$ = 0.1 T for the sample with $t_{IrMn}$ = 8 nm, showing current-induced switching of the FM layer. (h) $R_H$ vs. $B_z$ curves measured after the applied current pulses, demonstrating switching of the AFM interfacial spins.

The Hall bar samples were then subjected to a sequence of current pulses along the **x** direction, of varying amplitude $I_p$ and fixed width 50 ms, in a longitudinal applied field $B_x$ = 0.1 T [Fig. 1(b)]. Through the spin Hall effect (SHE), a charge current in the ± **x** direction should produce a spin polarization along the ± **y** direction for the positive spin-Hall angle of Pt [25]. The resulting spin current can switch the magnetization of PMA Co between the ± **z** directions, provided that both the current density and $B_x$ are large enough. Moreover, the absorption of transverse spin currents is found to vary with the FM thickness with a characteristic saturation length of 1.2 nm [26]. Thus in our devices, not only the 0.8 nm thick Co layer but also the AFM interfacial spins can be directly affected by SOT. Fig. 1(g) shows the measured $R_H$ after each current pulse for the sample with $t_{IrMn}$ = 8 nm, showing a square loop consistent with deterministic switching of the FM perpendicular magnetization.

$R_H$ vs. $B_z$ loops, obtained after the application of current pulses $I_p = \pm$ 26 mA in $B_x$ = 0.1 T, are shown in Fig. 1(h). The main part of the loop displays negative EB for $I_p$ = 26 mA (red) and positive EB for -26 mA (green). The opposite behaviors of $R_H$ vs. $B_z$ and $R_H$ vs. $I_p$ curves are observed for $B_x$ = -0.1 T (see Supplementary Fig. S1). The SOT induced EB switching is also found for $t_{IrMn}$ = 7 nm but not for $t_{IrMn}$ = 5 and 6 nm (see Supplementary Fig. S2).

Two-step hysteresis loops, similar to the one shown in Fig. 1(f), are commonly observed in as-deposited or zero-field cooled FM/AFM bilayers. They are related to the occurrence of a bi-domain state, in which the two domain populations are oppositely exchange biased due to opposite orientations of the uncompensated AFM spins at the FM/AFM interface [27,28]. The switching behavior observed in Fig. 1(h) is consistent with a change in the populations of the two domain types, due to a reorientation of interfacial AFM spins during the current pulse.

The effect of the Joule heating on the exchange bias reversal must be considered [20]. To estimate the temperature rise due to Joule heating, the resistance of the sample was measured during the current pulse for the sample with $t_{IrMn}$ = 7 nm. By comparing this to the measured temperature-dependence of resistance, a temperature rise of around 35 K was estimated for a 26 mA 50 ms current pulse (see Supplementary Fig. S3). In contrast, the blocking temperature for the $t_{IrMn}$ = 7 nm sample, defined as the temperature where the EB disappears, is around 450 K (see Supplementary Fig. S4). Therefore, we rule out a significant role of Joule heating in the observed switching.

We attribute the observed switching to the direct effect of the current-induced SOT on the uncompensated AFM spins at the FM/AFM interface. The spin current due to the SHE in the Pt layer induces a damping-like torque **m** × (**σ** × **m**) (along **y** direction) and a field-like torque **m** × **σ** (along **x** direction), where **m** is the interfacial spin moment and **σ** is the spin polarization of the spin current [29-35]. When the interfacial spins are deflected from the **z** direction due to SOT, switching into the direction of the FM layer magnetization will occur. The latter is determined by the relative alignments of $I_P$ and $B_x$ (Supplementary Fig. S1).

## IV. TUNING INTERFACIAL SPINS VIA SOT WITH LONGITUDINAL AND PERPENDICULAR FIELDS

Further investigations were focused on the tunability of AFM interfacial spins through SOT with the assistance of $B_x$ or $B_z$. Figure 2 shows $R_H$ vs. $B_z$ curves for the $t_{IrMn}$ = 8 nm sample after applying current pulses under different external fields, together with schematics of the interfacial spin configurations. For initial state, the observed two-step $R_H$ vs. $B_z$ switching behavior shows approximately equal weighting of its upper and lower parts, indicating no preference between upward and downward pinning directions for the interfacial spins [Fig. 2(a)]. We then investigated the effect of applying current pulses under different external magnetic field configurations, where the current pulse width was fixed at 50 ms and the magnitude of the current pulse ($I_P$) was fixed at 26 mA. The external magnetic field magnitude and direction during the current pulse is shown in the top panel of Fig. 2. After each treatment, the $R_H$ vs. $B_z$ curve (as shown in the middle panel of Fig. 2) was measured. After a positive current pulse of 26 mA under $B_x = 0.1$ T, a negative EB is observed, with the two-step $R_H$ vs. $B_z$ loop heavily weighted towards the lower part [Fig. 2(b)]. The opposite trend is found after applying - $I_P$ (-26 mA) under $B_x$ (0.1 T) [Fig. 2(c)]. Both curves are exhibited in Fig. 1(h). Further increases of $|I_P|$ (> 26 mA) or $B_x$ (> 0.1 T) do not further modify the $R_H$ vs. $B_z$ loops, indicating that the remaining oppositely aligned interfacial spins cannot be modified by SOT with the external field applied purely along **x**.

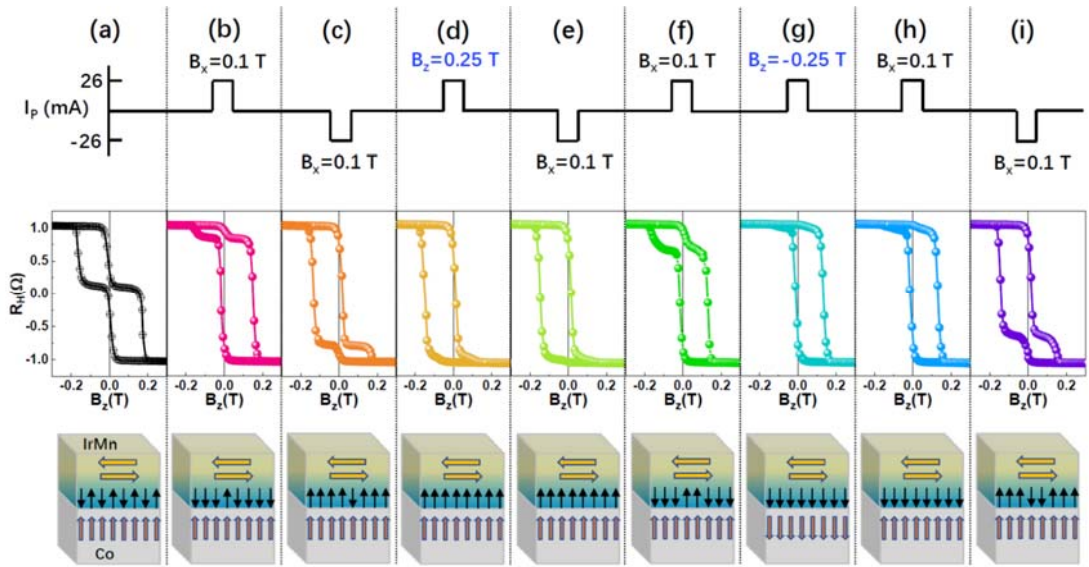

FIG. 2. AFM interfacial spins tuned by SOT. Sequences of current pulses with $B_x$ or $B_z$ applied to the sample with $t_{IrMn}$ = 8 nm, $R_H$ vs. $B_z$ curves, and schematics of the corresponding configurations of AFM and FM layers. (a) Initial state. (b) After applying $I_P$ = 26 mA in $B_x$ = 0.1 T. (c) After applying $I_P$ = -26 mA in $B_x$ = 0.1 T. (d) After applying $I_P$ = 26 mA in $B_z$ = 0.25 T. (e) After applying $I_P$ = -26 mA in $B_x$ = 0.1 T. (f) After applying $I_P$ = 26 mA in $B_x$ = 0.1 T. (g) After applying $I_P$ = 26 mA in -$B_z$ = -0.25 T. (h) After applying $I_P$ = 26 mA in $B_x$ = 0.1 T. (i) After applying $I_P$ = -26 mA in $B_x$ =

0.1 T.

Applying $I_p$ under $B_z = 0.25$ T results in a single-step $R_H$ vs. $B_z$ loop with positive EB [Fig. 2(d)], indicating a complete alignment of the interfacial spins in the direction of $B_z$. Subsequently applying $-I_p$ under $B_x = 0.1$ T does not affect the loop [Fig. 2(e)], while applying $+I_p$ under $B_x = 0.1$ T results in a partial switch [Fig. 2(f)]. Similarly, applying $I_p$ under $B_z = -0.25$ T results in a single-step loop with negative EB [Fig. 2(g)]. The opposite trend can then be seen in Fig. 2(h) and Fig. 2(i), as compared to that in Fig. 2(e) and Fig. 2(f), respectively. These results indicate that switching between multiple states of the AFM interfacial spins can be achieved via SOT combined with external magnetic fields.

## V. SYSTEMATIC VARIATION OF PULSE CURRENT AND MAGNETIC FIELDS

Next, we systematically investigate how pulse current intensity and the magnitude of the assisting magnetic fields affect the magnetic configuration of the HM/FM/AFM trilayer structure. As the $R_H$ vs. $I_p$ curves show in Fig. 3(a), the height of the loop $\Delta R_H = R_H^+ - R_H^-$ gradually increases on increasing the range of $I_p$ under fixed $B_x = 0.1$ T, saturating with $I_p \geq 22$ mA. Correspondingly the step in the $R_H$ vs. $B_z$ loops gradually moves to higher $R_H$ values [Fig. 3(b)]. The opposite direction of $I_p$ under the same $B_x$ induces the opposite shift of the magnetization step, as shown for the $I_P = -20$ mA loop in Fig. 3(b).

The shape of the $R_H$ vs. $B_z$ loop can be further controlled via SOT with varying $B_z$,

as shown in Fig. 3(c). Here, the initial state was set by applying $I_P$ = 22 mA under $B_z$ = -0.2 T to obtain a single-step loop. Subsequent pulses of $I_P$ = 22 mA under varying $B_z$ from 1 to 200 mT result in a continuously adjustable $R_H$ step height. The switched fraction, defined as the ratio of the $R_H$ step height to the saturation $R_H$ value, is plotted versus $B_z$ in Fig. 3(d). Two distinct behaviors are observed: the switched fraction increases sharply to ~82 % with $B_z$ from 1 to 5 mT (region **I**), and then gradually increases to 100 % with further increasing $B_z$ (region **II**). The curve's slope for region **I** is about two orders of magnitude higher than for region **II**. Significantly, the switched fraction of 82% marked by the dashed line in Fig. 3(d) at the boundary between regions **I** and **II** is close to that for $I_P \geq$ 22 mA with longitudinal field, seen in Fig. 3(b). Therefore, the SOT induced switching under $B_x$ is only effective up to the upper limit of region **I**.

Furthermore, we found that positive and negative $I_P$ have nearly the same effect on the switching under $B_z$ [see Supplementary Fig. S5(a)]. This is consistent with our interpretation of the switching as being due to the direct effect of the SOT on the AFM interfacial spins. A deflection of the interfacial spins from the perpendicular direction due to a current-induced SOT of either sign will enable their switching into the direction of $B_z$, in order to minimize the Zeeman energy. Accordingly, a smaller SOT (due to smaller pulsed current) requires a larger $B_z$ to flip the interfacial spins, as shown in Supplementary Fig. S6.

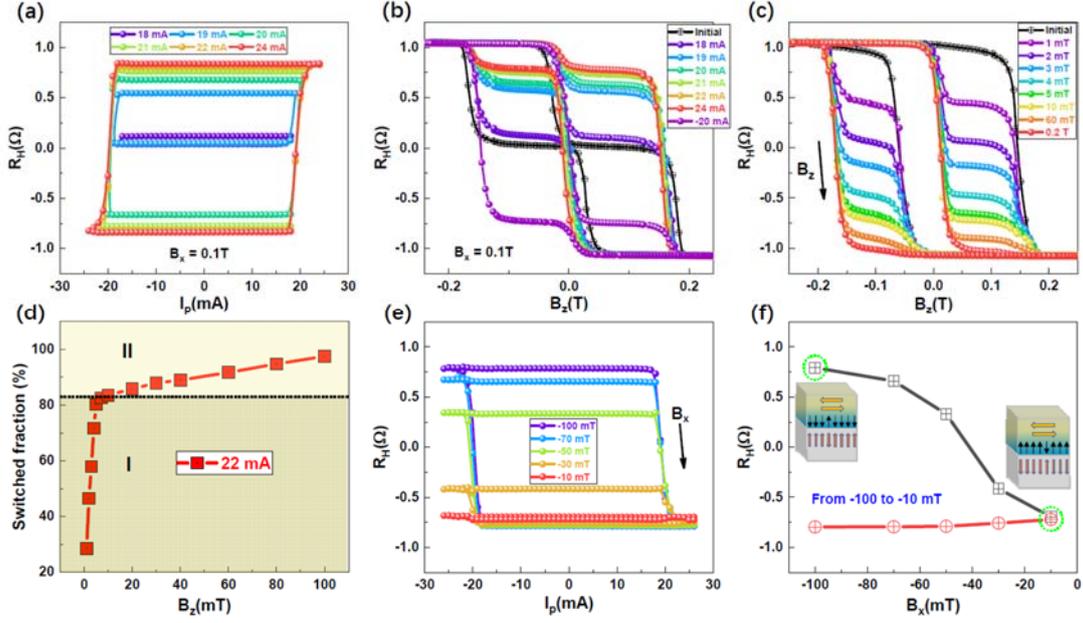

FIG. 3. Dependence on pulse current and magnetic field magnitudes. $R_H$ vs. $B_z$ and $R_H$ vs. $I_p$ curves for sample with $t_{IrMn}$ = 8 nm. (a) $R_H$ vs. $I_p$ curves with varying range of $I_p$ and fixed $B_x$ = 0.1 T. (b) $R_H$ vs. $B_z$ curves after varying $I_p$ and fixed $B_x$ = 0.1 T. (c) $R_H$ vs. $B_z$ curves after applying $I_p$ = 22 mA under varying $B_z$, starting from an initial state set by applying $I_p$ = 22 mA under $B_z$ = -0.2 T. (d) The switched fraction of interfacial spins obtained from the $R_H$ vs. $B_z$ curves in (c), as a function of $B_z$. Two distinct regions are observed (indicated as **I** and **II**) separated by the dashed line at ~82%, and the slope for region **I** is around two orders of magnitude larger than for region **II**. (e) $R_H$ vs. $I_p$ curves for varying $B_x$. (f) The upper ($R_H^+$, black) and lower ($R_H^-$, red) values of Hall resistance as a function of $B_x$ obtained from (e). The schematic configurations of AFM interfacial spins and FM layers at the points marked by green circles are illustrated.

The effect of SOT with varying $B_x$ on the interfacial spin configuration is also observed. As shown in Fig. 3(e) and Fig. 3(f), with changing $B_x$ from -100 to -10 mT, the $R_H^-$ stays nearly constant while the $R_H^+$ gradually reduces. Therefore, the SOT switching is gradually reduced with decreasing negative $B_x$. Similarly, the $R_H$ vs. $I_p$ curves with varying $B_x$ from 100 to 10 mT exhibit a constant $R_H^+$ and a gradual change of $R_H^-$ (see Supplementary Fig. S7). However, after annealing the sample in a magnetic field along **z**, the switching is found to be only weakly dependent on $B_x$ in the range 10-100 mT (see Supplementary Fig. S8), because the field-annealing induces an out-of-plane effective field which can assist the SOT switching. Therefore, the switching can take place in quite small $B_x$ in the field-annealed case.

**VI. ZERO-FIELD SOT INDUCED AFM INTERFACIAL SPIN SWITCHING**

We also observed a modification of the $R_H$ vs. $B_z$ loop induced by SOT in zero external field for the sample with $t_{IrMn}$ = 8nm (Fig. 4). The initial state was set by applying $I_P$ = 22 mA under $B_z$ = -0.2 T, and subsequent loops were obtaining after applying $I_P$ of varying magnitude under zero field. As shown in Fig. 4(a), after pulsing in zero field the $R_H$ vs. $B_z$ behavior transforms from a single-step loop similar to Fig. 2(g) for the initial state, to a two-step loop similar to Fig. 2(b). The effect is much more pronounced for the field-annealed sample [Fig. 4(b)]. Comparing the switched fractions versus $I_P$ in Fig. 4(c), a smaller threshold $I_P$ and a much larger switched fraction is observed for the field-annealed sample. The saturated state after zero-field SOT in Fig. 4(b) is close to the initial state of the as-deposited device [see Fig. 2(a)], with nearly equal upward and downward parts of the loop.

In HM/FM systems with PMA, it is necessary to break the symmetry between up and down magnetization directions in order to generate deterministic switching using SOT. Typically this is achieved by applying an in-plane magnetic field collinear with the electric current, but a lateral asymmetry [31], tilted magnetic anisotropy [36], anti-ferromagnetic layer [15], polarized ferroelectric substrate [33], interlayer exchange coupling [16,37], interfacial spin-orbit interaction [38] or competing spin currents [39] have also been introduced to achieve field-free deterministic switching. In our system, the field-free SOT induced interfacial spin switching should be related to the inequivalent upward and downward domain populations, which produces an effective out-of-plane field ($B_{z\text{-eff}}$). It can be considered as a training effect, in which the built-in $B_{z\text{-eff}}$ assists the SOT to switch the interfacial spins from a metastable single-domain state, to an equilibrium state with incomplete alignment of the interfacial spins.

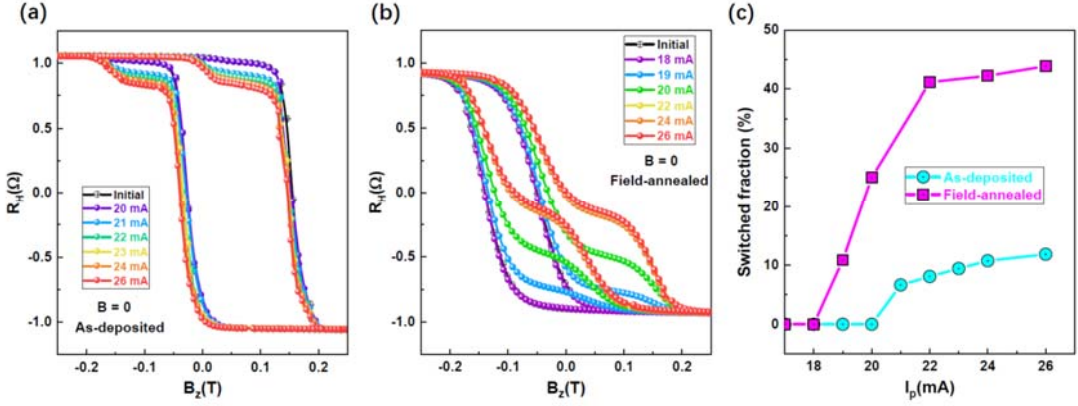

FIG. 4. SOT-induced switching under zero magnetic field. (a),(b) $R_H$ vs. $B_z$ curves after different pulsed currents for $t_{\text{IrMn}}$ = 8 nm in zero field. For (a) the initial state was set applying $I_P$ = 22 mA under $B_z$ = -0.2 T at room temperature. For (b) the initial state was set by annealing the sample at 250°C in a magnetic field $B_z$ = 0.7 T, and then field cooling to room temperature. (c) The switched fraction obtained from the $R_H$ vs. $B_z$ curves in (a) and (b), as a function of $I_p$ for the as-deposited and field-annealed devices.

## VII. DISCUSSIONS AND CONCLUSION

We have demonstrated a high controllability of the spin states at the FM/AFM interface via SOT. Multi-state switching is achieved using SOT in combination with external magnetic fields $B_z$ or $B_x$, while field-free switching from the fully aligned state was also realized. Our work provides a very efficient scheme for tuning of the uncompensated antiferromagnetic interfacial spin states via SOT, which will expand the designability of spintronic devices. For instance, the SOT-magnetic random access memory (MRAM) can potentially be realized by varying the FM/AFM interface via SOT, in contrast to the conventional design. Multiple resistance states and thus high density storage may be achieved in this SOT-MRAM cell. Furthermore, combining with the conventional field-annealing and the pulsed electrical currents approaches will open up more potential applications in spintronic devices. For example, if the EB is initially set along a preferred in-plane axis (**x** or **y**) by field-annealing, the current pulses can induce EB in perpendicular direction without disturbing the in-plane EB. For magnetic sensors containing many cells with different exchange bias directions, the current-pulse offers a convenient approach to tune the EB in different directions, respectively. In addition, the precise control of interfacial spins at a FM/AFM interface by SOT might result in a multi-state perpendicular ferromagnet, which has a potential application in a synaptic emulator for neuromorphic computing.

## ACKNOWLEDGEMENTS

This work was supported by the National Key R&D Program of China No.s 2017YFA0303400 and 2017YFB0405700. This work was also supported by the NSFC Grant No.s 11474272 and 61774144. The Project was sponsored by the Chinese Academy of Sciences, Grant No.s QYZDY-SSW-JSC020, XDB28000000, and XDPB12 as well.

# Tuning interfacial spins in antiferromagnetic / ferromagnetic / heavy metal heterostructures via spin-orbit torque

X. H. Liu[1,2], K. W. Edmonds[3], Z. P. Zhou[1,2], K. Y. Wang[1,2,4,5*]

[1]*State Key Laboratory for Superlattices and Microstructures, Institute of Semiconductors, Chinese Academy of Sciences, Beijing 100083, China*

[2] *Center of Materials Science and Optoelectronics Engineering, University of Chinese Academy of Sciences, Beijing 100049, China*

[3] *School of Physics and Astronomy, University of Nottingham, Nottingham NG7 2RD, United Kingdom*

[4] *Beijing Academy of Quantum Information Sciences, Beijing 100193, China*

[5]*Center for Excellence in Topological Quantum Computation, University of Chinese Academy of Science, Beijing 100049, China*

\* Corresponding e-mail: kywang@semi.ac.cn

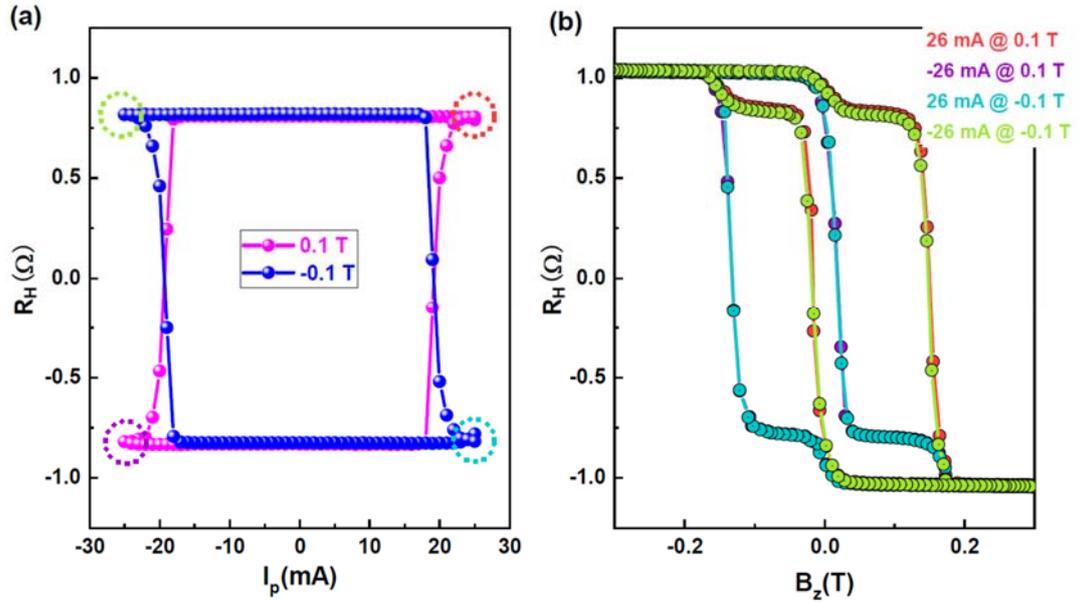

FIG. S1. $R_H$ vs. $I_p$ and $R_H$ vs. $B_z$ curves for $t_{IrMn}$ = 8 nm, for positive and negative in-plane fields $B_x$. (a) $R_H$ vs. $I_p$ curves at $B_x$ = 0.1 T and -0.1 T. (b) $R_H$ vs. $B_z$ curves for states after applying different current pulses and $B_x$ (as indicated by different color circles). The opposite behaviors of $R_H$ vs. $B_z$ and $R_H$ vs. $I_p$ curves are observed for $B_x$ = 0.1 T and -0.1 T.

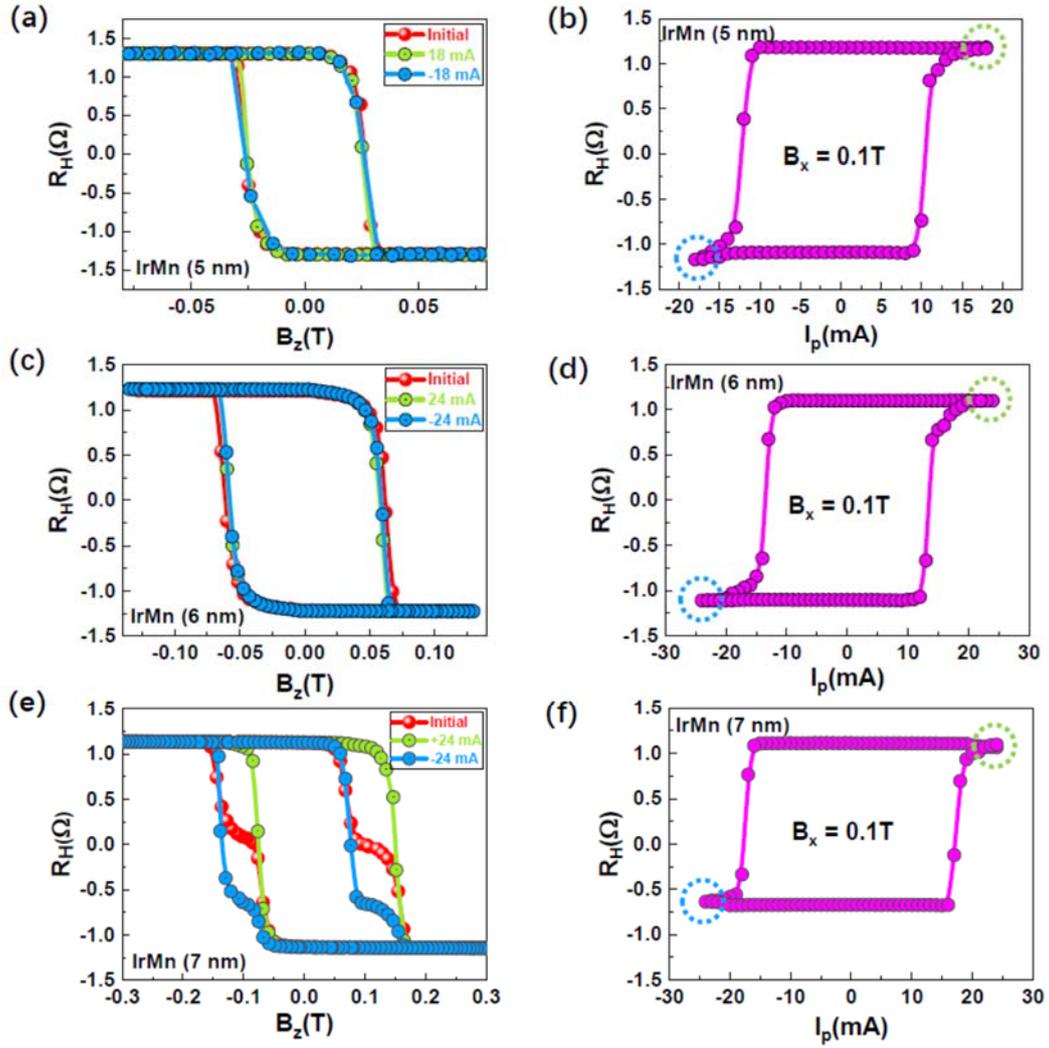

FIG. S2. $R_H$ vs. $B_z$ and $R_H$ vs. $I_p$ curves for samples with $t_{IrMn}$ = 5 nm (a) and (b), $t_{IrMn}$ = 6 nm (c) and (d), and $t_{IrMn}$ = 7 nm (e) and (f), respectively. The magnetization switching by SOT in $B_x$ = 0.1 T can be observed for all the samples, while the switching of exchange bias is not observed for 5 and 6 nm samples but is observed for 7 nm sample after applying large enough pulsed currents (as different color circles indicate).

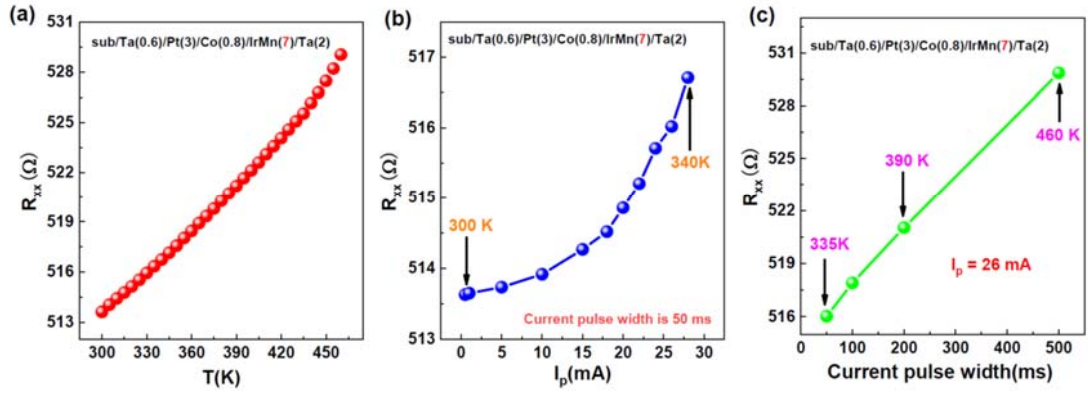

FIG. S3. Determination of the temperature increase due to Joule heating during the current pulse. (a) Resistance versus temperature. (b) Resistance measured during the current pulse, versus the current pulse magnitude $I_p$ for fixed pulse width of 50 ms. (c) Resistance measured during the current pulse, versus the pulse width for fixed magnitude $I_p$ = 26 mA.

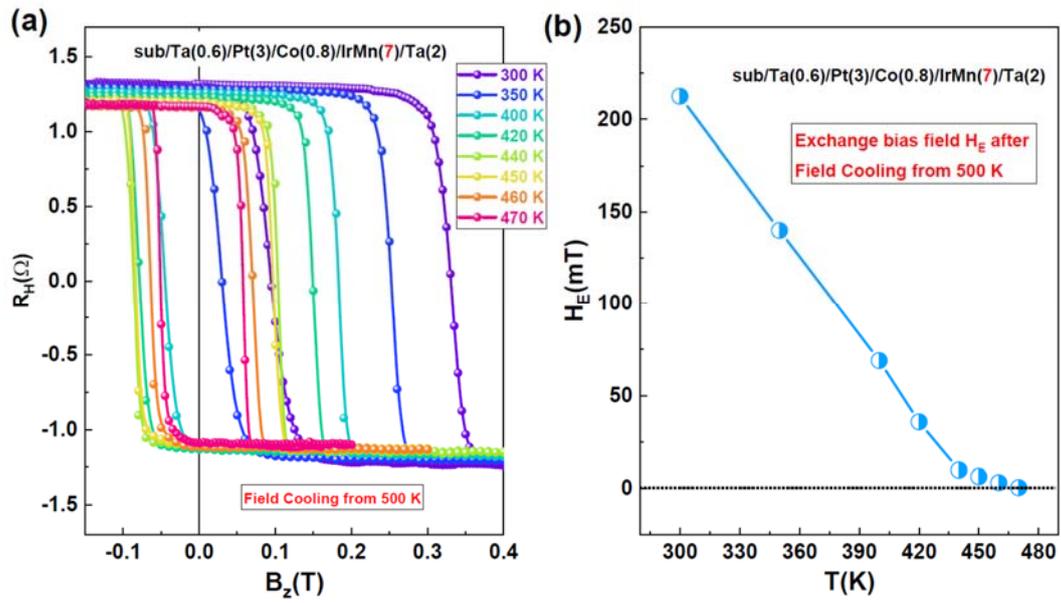

FIG. S4. Determination of the blocking temperature of the IrMn layer. (a) $R_H$ vs. $B_z$ loops at different temperatures, after field-cooling from 500 K. (b) Exchange bias field versus temperature extracted from the data in (a).

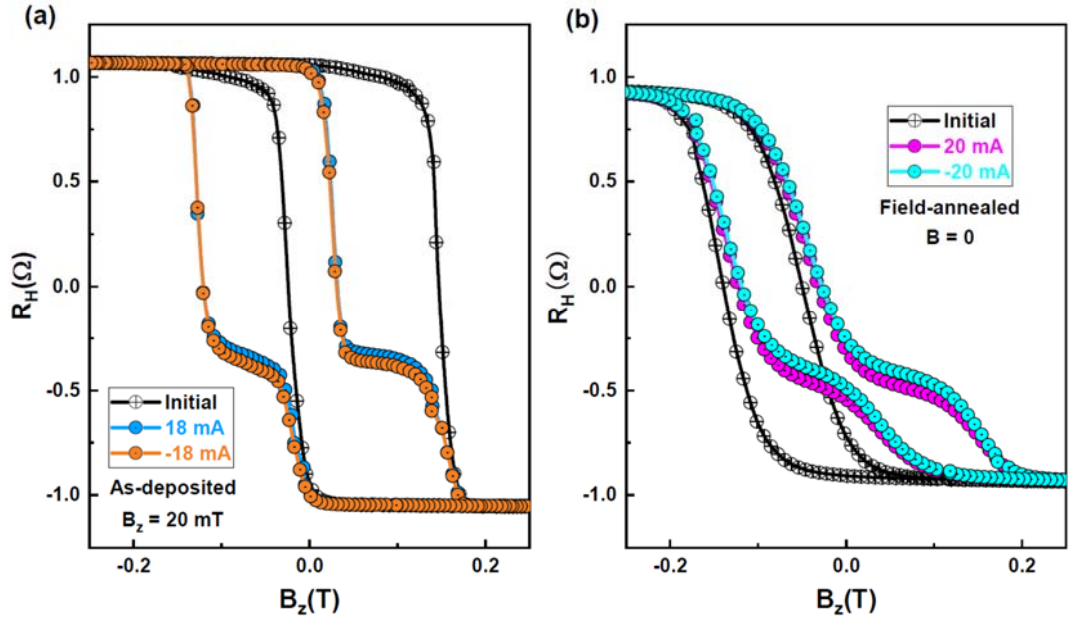

FIG. S5. $R_H$ vs. $B_z$ curves in the fully-aligned state, and after applying positive and pulsed currents. The results show a nearly identical effect of positive and negative SOT. (a) As-deposited sample with $t_{IrMn}$ = 8 nm, where the initial state was set utilizing 22 mA under $B_z$ = -200 mT. (b) Sample with $t_{IrMn}$ = 8 nm after annealing at 250 °C in $B_z$ = 0.7 T and field-cooling. In (a) a field of $B_z$ = 20 mT was applied during pulsing, while in (b) the pulsing is performed in zero field.

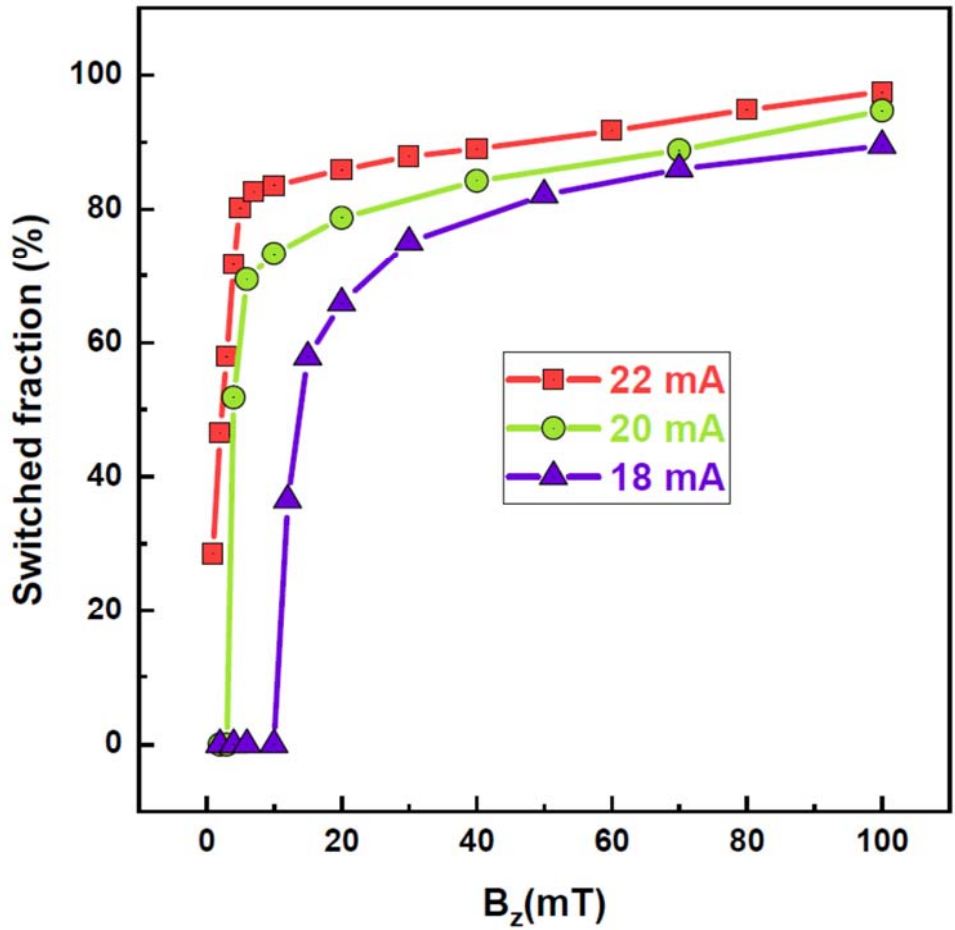

FIG. S6. Switched fraction of interfacial spins determined from the step in $R_H$ vs. $B_z$ curves, as a function of perpendicular field $B_z$ for different pulsed currents. With decreasing current pulse magnitude, a larger $B_z$ must be applied to achieve switching of the interfacial spin state. For $I_p \leq 20$ mA, a threshold field must be applied for switching to occur, which reaches around 12 mT for 18 mA.

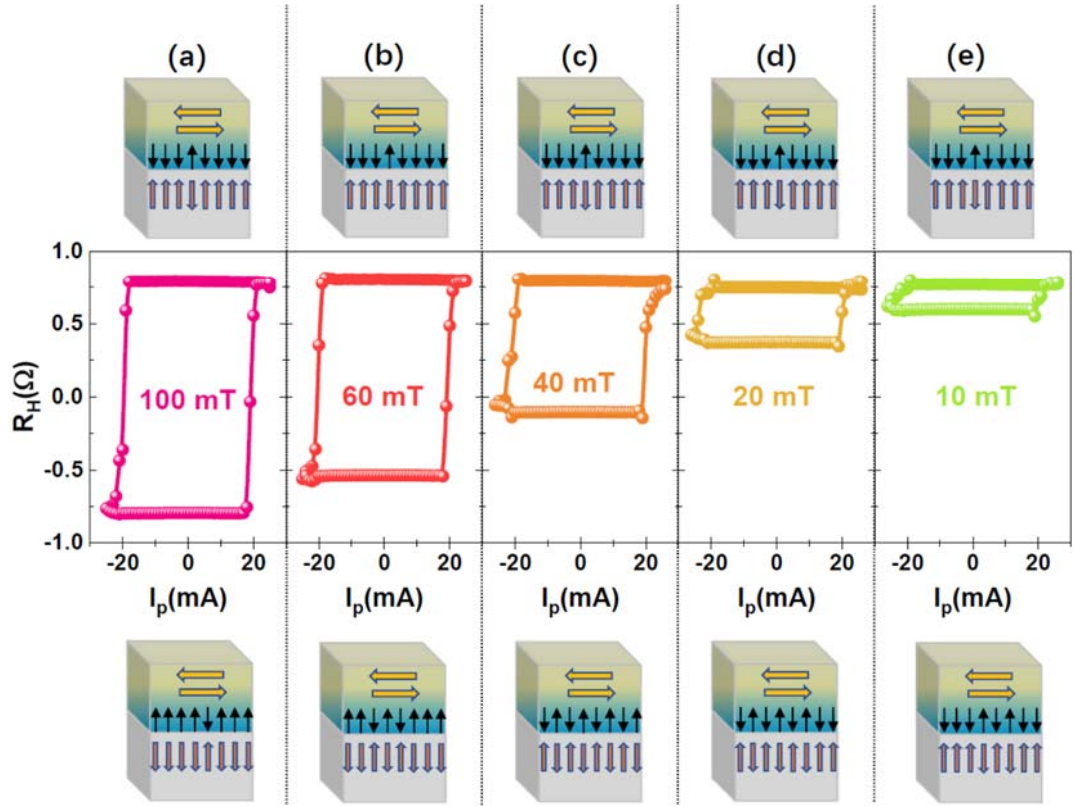

FIG. S7. $R_H$ vs. $I_p$ curves and schematic configurations of AFM and FM layers at $R_H^+$ and $R_H^-$ for the sample with $t_{IrMn}$ = 8 nm, for varying $B_x$ from 100 to 10 mT (a-e). In this process, the $R_H^+$ stays constant while the $R_H^-$ gradually moves closer to $R_H^+$, opposite to the behavior shown in Fig. 3(e) for negative $B_x$.

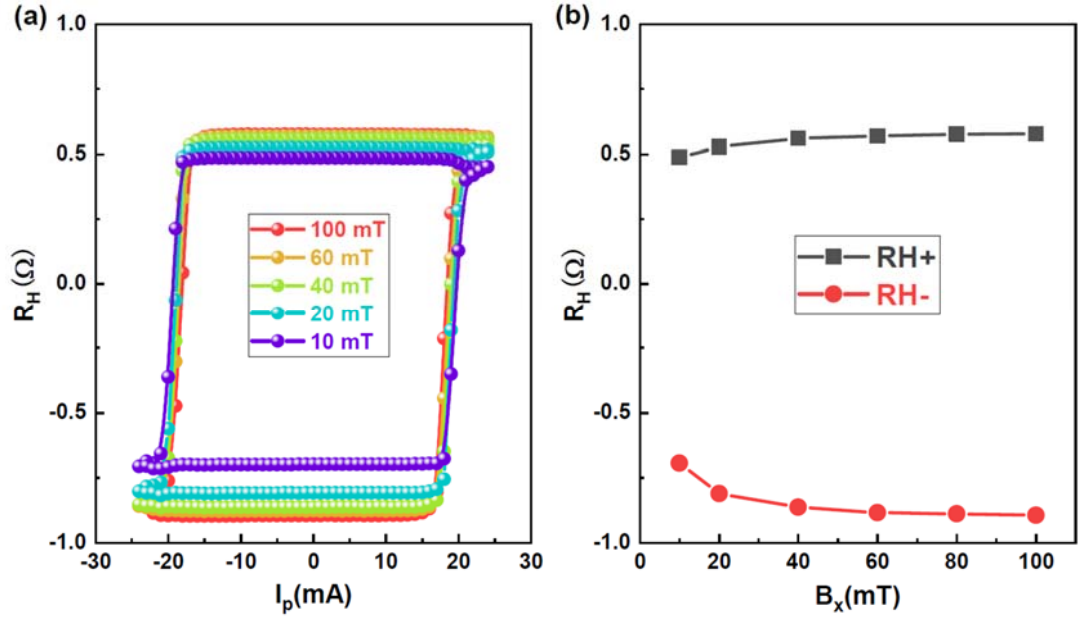

FIG. S8. (a) $R_H$ vs. $I_p$ curves for the sample with $t_{IrMn}$ = 8 nm after field-annealing with field along **z**, for varying $B_x$ from 100 to 10 mT applied during the pulse. Compared to the as-deposited sample shown in Fig. S7, both the $R_H^+$ and the $R_H^-$ vary only slightly with varying $B_x$. The $R_H^+$ and $R_H^-$ as a function of $B_x$ are summarized in (b). The field-annealing treatment results in a built-in effective magnetic field due to the AFM interfacial spins, enabling deterministic SOT switching of the FM layer under small $B_x$.